\documentclass[aps,prl,twocolumn,groupedaddress]{revtex4}

\usepackage{graphicx}

\begin{document}

\renewcommand{\vec}[1]{{\mathbf #1}}

\title{Quantum theory of dynamic multiple light scattering}
\author{S.E. Skipetrov}
\email[]{Sergey.Skipetrov@grenoble.cnrs.fr}
\affiliation{Laboratoire de Physique et Mod\'elisation des Milieux Condens\'es\\
CNRS and Universit\'e Joseph Fourier, 38042 Grenoble, France}

\date{\today}

\begin{abstract}
We formulate a quantum theory of dynamic multiple light scattering in fluctuating disordered media
and calculate the fluctuation and the autocorrelation function of photon number operator for light
transmitted through a disordered slab.
The effect of disorder on the information capacity of a quantum communication channel
operating in a disordered environment is estimated and the use of squeezed light in diffusing-wave spectroscopy
is discussed.
\end{abstract}


\maketitle


Many disordered media (suspensions, emulsions, foams, semiconductor powders, biological tissues, etc.) scatter light diffusely and appear opaque.
Statistical properties of light scattered from disordered media are of fundamental
importance for such applications as wireless communications \cite{moustakas00,andrews01,skip03} and different types of diffuse optical spectroscopies (diffuse transmission spectroscopy \cite{kaplan94}, diffusing-wave spectroscopy \cite{weitz93,maret97}, etc.).
With a few notable exceptions \cite{patra99,lodahl05}, most of theoretical studies in this field of research have been conducted by treating light as a classical wave \cite{rossum99,akker04}, without accounting for
the quantum nature of light.
Meanwhile, recent experiments demonstrate that quantum properties of light
in disordered media (such as quantum noise \cite{lodahl05exp} and photon statistics \cite{balog06})
can be of interest from both fundamental and applied points of view. Moreover,
the progress in generation of light with nonclassical properties (single photons and light in number states \cite{lounis05}, squeezed light \cite{breit97}, entangled photons \cite{benson00}) suggests that sources of nonclassical light could soon become accessible to a wide scientific community. This calls for a better theoretical understanding of
quantum aspects of multiple light scattering.

The purpose of the present Letter is to fill up an important gap in the quantum theory of multiple light scattering
by developing a quantum description of \textit{dynamic} multiple light scattering in fluctuating
disordered media.
In contrast to the previous work \cite{patra99,lodahl05}, we take into account the evolution of
disordered medium (e.g., Brownian or more complex motion of scattering centers) \textit{in the course} of
a multiple scattering experiment. This leads to an interesting and reach problem of quantum
dynamics, where the quantum fluctuations of light are coupled with the classical fluctuations of the medium.
Our results allow us to estimate
the impact of disorder on the information capacity of quantum communication channels
and to lay a foundation for the use of nonclassical light as a probe of disordered media.

\begin{figure}
\includegraphics[width=7.0cm,angle=0]{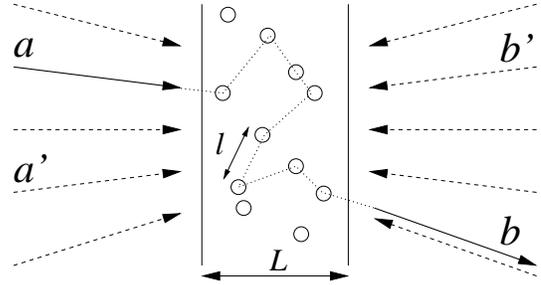}
\caption{\label{fig1}
An ensemble of mobile scattering particles (circles) is illuminated from the left by a wave in the
incoming mode $a$.
All other incoming modes are in vacuum states (modes $a^{\prime}$ on the left and
$b^{\prime}$ on the right).
We are interested in the fluctuations of transmitted light in the outgoing mode $b$.
Dotted line illustrates the multiple scattering with a mean free path $l \ll L$ that light undergoes inside the medium.}
\end{figure}

We consider a slab of disordered medium that, for concreteness, we assume to be a suspension of mobile,
elastically scattering particles (see Fig.\ \ref{fig1}).
For each frequency $\omega$, light incident on the slab can be decomposed in $M$ transverse modes that we denote by indices $a^{\prime}$ and $b^{\prime}$ for modes on the left and on the right of the slab, respectively. All but one of the incoming modes are assumed to be in vacuum states. The state of the mode $a^{\prime} = a$ on the left can be arbitrary and it corresponds to the light that one sends into the medium in an experiment. We are interested in the properties of multiply scattered light in the transmitted, outgoing mode $b$.

To quantize the electromagnetic field we use the same method as Beenakker {\em et al.} \cite{patra99} and Lodahl {\em et al.} \cite{lodahl05} who considered disordered media with immobile scattering centers.
An annihilation operator ${\hat a}_i(\omega)$ is associated with each mode and standard
bosonic commutation relations are assumed \cite{patra99,lodahl05}:
\begin{eqnarray}
[ {\hat a}_{i}(\omega), {\hat a}_{j}^{\dagger}(\omega^{\prime}) ] =
\delta_{ij} \delta(\omega-\omega^{\prime}),
\label{comrel}
\end{eqnarray}
where $i$ and $j$ run either over all incoming modes or over all outgoing modes. We now note that, as suggested by Fig.\ \ref{fig1}, the fluctuating disordered medium acts as a beam splitter having $M$ input and $M$ output ports and time-dependent transmission and reflection coefficients
$\tau_{a^{\prime} b}(t)$ and $\rho_{b^{\prime} b}(t)$.
By analogy with the work of Hussain {\em et al.} \cite{hussain92} who dealt with a `regular' beam splitter ($M = 2$) having time-dependent properties, we assume that $\tau_{a^{\prime} b}(t)$ and $\rho_{b^{\prime} b}(t)$ are frequency-independent and that they do not vary significantly during the typical time it takes for light to cross the disordered sample. The former assumption requires a sufficiently narrow bandwidth $B$ of the incident light, whereas the latter one is justified provided that the dynamics of the medium is slow. With these assumptions in hand, the input-output relations for ${\hat a}_i(\omega)$ read
\begin{eqnarray}
{\hat a}_b(t) = \sum\limits_{a^{\prime}} \tau_{a^{\prime} b}(t) {\hat a}_{a^{\prime}}(t) +
\sum\limits_{b^{\prime}} \rho_{b^{\prime} b}(t) {\hat a}_{b^{\prime}}(t),
\label{iorelations}
\end{eqnarray}
where ${\hat a}_i(t)$ denotes a Fourier transform of ${\hat a}_i(\omega)$.
The use of ${\hat a}_i(t)$ is permissible provided that $B$ is much smaller than the central frequency $\omega_0$ of the incident light \cite{hussain92,blow90}.
Similarly to the case of immobile scatterers \cite{patra99,lodahl05}, Eq.\ (\ref{iorelations}) conserves commutation relations (\ref{comrel}), which is due to the unitarity of the scattering matrix composed of $\tau_{a^{\prime} b}(t)$ and $\rho_{b^{\prime} b}(t)$ in the absence of absorption. However, in contrast to the previous work \cite{patra99,lodahl05}, Eq.\ (\ref{iorelations}) implies a possibility of energy exchange between different spectral components upon transmission of light through a fluctuating disordered medium. This becomes clear when we back Fourier-transform Eq.\ (\ref{iorelations}) because products of time-dependent quantities become convolutions in the frequency space. Such an energy exchange is by no means a nonlinear phenomenon, but is a manifestation of Doppler effect due to scattering of light on moving particles.

To establish a link between ${\hat a}_i(t)$ and measurable quantities, we define a flux operator
${\hat I}_i(t) = {\hat a}_i^{\dagger}(t) {\hat a}_i(t)$ that, when integrated from $t$ to $t + T$, yields a photon number operator ${\hat n}_{i}(t, T)$ corresponding to the number of photocounts measured by an ideal, fast photodetector illuminated by light in the mode $i$ during some sampling time $T$ \cite{blow90}.
A relation between the average values of operators ${\hat n}_b(t, T)$ and  ${\hat n}_a(t, T)$ corresponding to the outgoing mode $b$ and the incoming mode $a$, respectively, is readily found from Eqs.\ (\ref{comrel}) and (\ref{iorelations}):
$\overline{\langle {\hat n}_b \rangle} =
{\overline T}_{ab} \langle {\hat n}_a \rangle$. Here $T_{ab}(t) = \left| \tau_{ab}(t) \right|^2$,
the angular brackets denote the quantum mechanical expectation value, and the bar denotes averaging over disorder. By the ergodicity hypothesis, the simultaneous averaging $\overline{\langle \ldots \rangle}$ over both disorder and quantum fluctuations can be replaced by time averaging in a realistic experiment with a fluctuating disordered medium, provided that the incident light is stationary.
On the contrary, there is no simple way of obtaining averages over either disorder or quantum fluctuations separately.

Although the relation between $\overline{\langle {\hat n}_b \rangle}$ and $\langle {\hat n}_a \rangle$ is independent of the quantum state of the incident light, the latter is crucial for the second order statistics of the photon number operator.
To illustrate this, we consider a particularly simple case of a stationary, quasi-monochromatic incident beam for which both
$\langle {\hat I}_a(t) \rangle$ and
$\langle :{\hat I}_a(t_1) {\hat I}_a(t_2): \rangle$, with $: \ldots :$ denoting normal ordering \cite{mandel95},
can be considered time-independent on the scale of $T$.
This includes a number of practically important examples on which we will dwell in more detail:
coherent light, continuous-wave quadrature squeezed light with a rectangular squeezing spectrum of width $B \ll 1/T$ around the central frequency of a perfectly monochromatic mean field $A_0 \exp(-i \omega_0 t + i \phi)$ \cite{zhu90},
light in a generalized continuum number state $|n, \xi \rangle$ with a box-shaped envelope function $\xi(t) = (t_{\mathrm{max}}-t_{\mathrm{min}})^{-1/2} \exp(-i \omega_0 t)$, where $t_{\mathrm{min}} \leq t_1, t_2 \leq t_{\mathrm{max}}$ \cite{fnote},
and chaotic light with a long coherence time $t_{\mathrm{coh}} \gg T$.
Making use of Eqs.\  (\ref{comrel}) and (\ref{iorelations}), we
find that the normalized variance of photon number fluctuations in the outgoing mode $b$ can be written as
\begin{eqnarray}
\delta_{b}^2 &=&  \frac{\mathrm{Var}[{\hat n}_b]}{\overline{\langle {\hat n}_b \rangle}^2}
=
\nonumber \\
&=& \frac{1}{\overline{\langle {\hat n}_b \rangle}} + \delta_{\mathrm{class}}^2 +
\frac{Q_a}{\langle {\hat n}_a \rangle} \left(1 + \delta_{\mathrm{class}}^2 \right).
\label{db}
\end{eqnarray}
Here
$\mathrm{Var}[{\hat n}_b] = \overline{\langle {\hat n}_b^2 \rangle} - \overline{\langle {\hat n}_b \rangle}^2$,
$\langle {\hat n}_a \rangle$ is the average photon number and 
$Q_a =  \mathrm{Var}[{\hat n}_a]/\langle {\hat n}_a \rangle - 1$
is the Mandel parameter \cite{mandel95} corresponding to the incident light beam.
$\delta_{\mathrm{class}}^2 =
(2/T) \int_0^{T} \mathrm{d} t (1 - t/T) C_T^{ab}(t)$ is the normalized variance of time-integrated intensity in the mode $b$ that one obtains from Eq.\ (\ref{iorelations}) by ignoring the quantum nature of light and dropping the operator `hats' of ${\hat a}_i(t)$; 
$C_T^{ab}(t) = \overline{T_{ab}(t_1) T_{ab}(t_1 + t)}/{\overline{T}_{ab}^2} - 1$ is the autocorrelation function of $T_{ab}(t)$.
$\delta_{\mathrm{class}}^2$ represents the `classical' part of $\delta_{b}^2$ in the sense that it is determined uniquely by the (classical) fluctuations of transmission coefficient $T_{ab}(t)$ due to the motion of scattering centers in the medium.

The main conclusion following from Eq.\ (\ref{db}) is that, in the general case, the variance of photon number fluctuations in transmission through a fluctuating disordered medium is not simply a sum of quantum and classical (due to
disorder) contributions, but rather a much more complicated object.
In particular, the last term of Eq.\ (\ref{db}) mixes classical fluctuations and quantum noise in a multiplicative way.
Only for incident light in the coherent state $Q_a = 0$ and Eq.\ (\ref{db}) reduces to a sum of shot noise $1/\overline{\langle {\hat n}_b \rangle}$ and classical noise $\delta_{\mathrm{class}}^2$. This is the minimum value of $\delta_b^2$ that can be obtained for light in a state admitting classical description. An `excess noise' $\delta_{\mathrm{excess}}^2$ --- defined as a difference between $\delta_b^2$ and $1/\overline{\langle {\hat n}_b \rangle} + \delta_{\mathrm{class}}^2$ --- arises for other states. For chaotic light, for example, $Q_a = \langle {\hat n}_a \rangle$ and $\delta_{\mathrm{excess}}^2 = 1 +\delta_{\mathrm{class}}^2$. Note that for both coherent and chaotic states we do not really need the quantum model developed here, but can instead use the semiclassical Mandel's formula \cite{mandel95} as it has been done in Ref.\ \cite{balog06}. 

\begin{figure}
\includegraphics[height=7.0cm,angle=0]{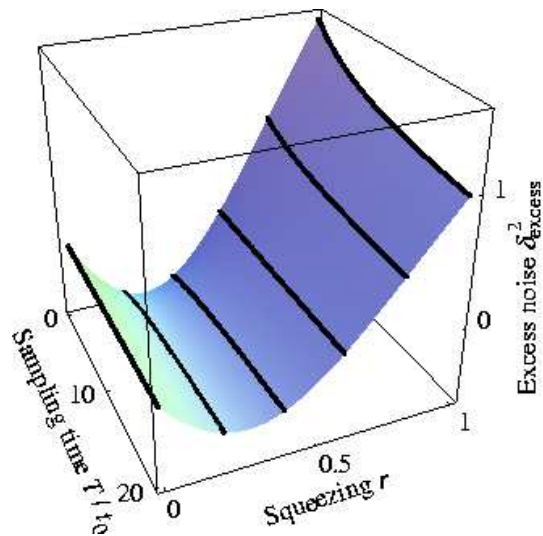}
\caption{\label{fig2}
Excess noise of photon number ${\hat n}_b$ in the transmitted mode $b$ for the incident light (mode $a$) in the continuous-wave amplitude-squeezed state with a narrow rectangular squeezing spectrum of width $B \ll 1/T$.
The excess noise is defined as a difference between the variance of photon number fluctuations
$\delta_b^2$, the shot noise $1/\overline{\langle {\hat n}_b \rangle}$ and the classical noise $\delta_{\mathrm{class}}^2$.
It is shown as a function of sampling time $T$ and squeezing parameter $r$ for a suspension
of small dielectric particles in Brownian motion (diffusion coefficient $D_B$) as depicted in Fig.\ \ref{fig1}.
The unit of time is $t_0 = l^2/6 k^2 D_B L^2$, where
$l$ is the mean free path, $L \gg l$ is the sample thickness, and $k \gg 1/l$ is the wave number of light.
The intensity carried by the mean field $A_0^2$ is set equal to $B$. 
Thick black lines show the excess noise as a function of sampling time for $r = 0$, 0.2, 0.4, 0.6, 0.8 and 1.0.}
\end{figure}

We now turn to nonclassical light --- light in states that cannot be described classically and to which the Mandel's formula does not apply. For incident light in the number state $|n_a, \xi \rangle$, $Q_a = -1$ and the excess noise is {\em negative}: the fluctuations of transmitted photon number are suppressed below the simple sum of shot noise $1/{\overline{\langle {\hat n}_b \rangle}}$ and classical fluctuation $\delta_{\mathrm{class}}^2$. $\delta_{b}^2$ becomes even less than the bare shot noise $1/{\overline{\langle {\hat n}_b \rangle}}$ if $n_a < 1 + \delta_{\mathrm{class}}^{-2}$. An interesting situation occurs for the single-photon number state ($n_a = 1$): the relative fluctuation of transmitted photon number becomes independent of $\delta_{\mathrm{class}}^2$. Another important example of nonclassical light is the continuous-wave squeezed light for which 
$Q_a = \langle {\hat n}_a \rangle f(r, \phi, A_0^2/B, 1)$ \cite{zhu90}, where
$r$ is the squeezing parameter and we define an auxiliary function
$f(r, \phi, x, y) =
[ x y ( e^{-2 r}\cos^2 \phi + e^{2 r}\sin^2 \phi - 1 )
+ y^2 \sinh^2 r \cosh 2r ]/
( x + \sinh^2 r )^2$. For the amplitude-squeezed light ($\phi = 0$) the Mandel parameter is negative if  $r < r_0$, with $r_0$ depending on $A_0^2/B$, leading to $\delta_{\mathrm{excess}}^2 < 0$, similarly to the number state. This is illustrated in Fig.\ \ref{fig2} for the case of $C_T^{ab}(t) = (t/t_0)/\sinh^2\sqrt{t/t_0}$ which corresponds to a suspension of small Brownian particles \cite{weitz93,maret97} (the characteristic decay time $t_0$ is defined in the caption of Fig.\ \ref{fig2}). With $A_0^2/B = 1$ chosen for this figure the excess noise is negative for $r < r_0 \simeq 0.6$.

Equation (\ref{db}) can be used to estimate the effect of
disorder on the information capacity of quantum communication channels operating in disordered environments. As an example, let us consider a communication channel connecting Alice on the left of the disordered medium in Fig.\ \ref{fig1} and Bob on the right. Alice communicates with Bob by sending a sequence of number states, each of duration $T$ and containing up to $n_a^{\mathrm{max}}$ photons, in the incoming mode $a$. Bob reconstructs the message of Alice by counting photons in the outgoing mode $b$ during the same time intervals $T$.
By using the theory of linear bosonic communication channels \cite{caves94} and Eq.\ (\ref{db}),
we find that the information capacity of such a channel is
\begin{eqnarray}
{\cal C} &=& \frac{n_a^{\mathrm{max}} \overline{T}_{ab}}{T e \ln 2}
\left\{
1 + \frac{1}{2} n_a^{\mathrm{max}} \overline{T}_{ab} \right.
\nonumber \\
&\times& \left. \left[ \frac{1}{n_a^{\mathrm{max}}} - \frac{1}{e} - \delta_{\mathrm{class}}^2 \left( 1 - \frac{1}{n_a^{\mathrm{max}}} \right) \right] \right\} \;\frac{\mbox{bits}}{\mbox{s}},
\label{cap}
\end{eqnarray}
where we drop higher-order terms in $n_a^{\mathrm{max}} \overline{T}_{ab} \ll 1$, which is justified by the condition $\overline{T}_{ab} \ll 1$ typical for transmission of light through a diffusely scattering medium.
As could be expected, the (classical) fluctuations $\delta_{\mathrm{class}}^2$ reduce the capacity of the channel. However, this reduction is only second-order in $n_a^{\mathrm{max}} \overline{T}_{ab} \ll 1$. The drop of capacity in the presence of disorder is therefore mainly due to the reduced average value of transmission coefficient $T_{ab}$ rather than to the fluctuations of the latter. Interestingly, ${\cal C}$ appears to be independent of $\delta_{\mathrm{class}}^2$ if $n_a^{\mathrm{max}} = 1$ (Alice sends either 1 or no photons).

\begin{figure}
\includegraphics[height=7.0cm,angle=0]{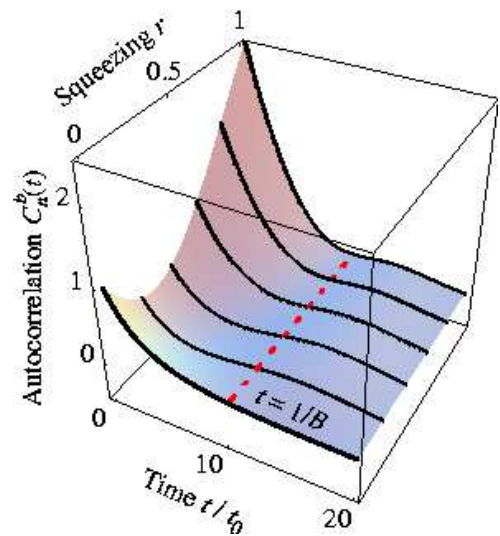}
\caption{\label{fig3}
Autocorrelation function of photon number fluctuations in transmission of amplitude-squeezed light through a suspension of small dielectric particles in Brownian motion for the same values of parameters as in Fig.\ \ref{fig2} and $t_0 = 0.1/B$.
The dashed red line illustrates that $C_n^b(t)$ is independent of $r$ for $t = 1/B$.}
\end{figure}

Correlation of photon numbers in the same mode $b$ but in two different, nonoverlapping time intervals,
\begin{eqnarray}
C_{n}^b(t) = \frac{\overline{\langle {\hat n}_b(t_1, T) {\hat n}_b(t_1 + t, T)\rangle}}{
\overline{\langle {\hat n}_b(t_1, T) \rangle} \;
\overline{\langle {\hat n}_b(t_1 + t, T) \rangle}} - 1,
\label{cf0}
\end{eqnarray}
is also affected by the quantum state of incident light and can be reduced to
\begin{eqnarray}
C_{n}^b(t) = \left[ 1 + C_{n}^a(t) \right] C_T^{ab}(t) +  C_{n}^a(t),
\label{cf}
\end{eqnarray}
where $C_{n}^a(t)$ is the autocorrelation function of photon number fluctuations in the incident mode $a$.
In order to put Eq.\ (\ref{cf}) in a simple form, the typical decay time of $C_T^{ab}(t)$ is assumed to be much longer than $T$.
Equation (\ref{cf}) is more general than Eq.\ (\ref{db}) and requires
neither stationarity nor monochromaticity of incident light.
For light in the coherent state $C_n^a(t) = 0$ and $C_n^b(t) = C_T^{ab}(t)$. This result actually lies at the heart of diffusing-wave spectroscopy (DWS) and other photon correlation techniques that identify the measured $C_n^b(t)$ with $C_{T}^{ab}(t)$ and interpret it in terms of classical wave scattering.

For nonclassical states of incident light $C_n^b(t)$ is not equal to $C_T^{ab}(t)$.
For the number state $|n_a, \xi \rangle$, $C_n^a(t) = -1/n_a$ and $C_n^b(t)$ can take negative values and becomes independent of time for $n_a = 1$.
These strange properties are direct consequences of the conservation of total number of photons --- a conservation
law which does not apply to the coherent state where the number of photons is not a good quantum number.
For squeezed light we find $C_n^a(t) = f[r, \phi, A_0^2/B, \mathrm{sinc}(\pi B t)]$. As illustrated in Fig.\ \ref{fig3}, squeezing induces oscillations of $C_n^b(t)$ and decreases its value at short $t$. The latter reaches a minimum at some $r = r_1$ ($r_1 \simeq 0.3$ in Fig.\ \ref{fig3}) and then grows again, exceeding 1 at large $r$.
The results presented in Fig.\ \ref{fig3} open a way to performing DWS with squeezed light; the necessary experimental setup is completely analogous to the `traditional' DWS one \cite{weitz93,maret97}, except that a source of squeezed light should be used instead of a laser.

Provided that $T_{ab}(t)$ is replaced by
$T_a(t) = \sum_b T_{ab}(t)$ in the definition of $C_T^{ab}(t)$,
the main results of this Letter --- Eqs.\ (\ref{db}) and (\ref{cf}) --- also hold for the variance
and the autocorrelation function of ${\hat n} = \sum_b {\hat n}_b$ --- a photon number operator corresponding to the `total transmission' measurement. Equations (\ref{db}) and (\ref{cf}) thus implicitly
account for all possible long-range
(in space or in time) correlations of $T_{ab}(t)$ that were extensively studied previously \cite{rossum99,akker04}. 
The effect of long-range correlations on photon number fluctuations in total transmission through a
suspension of scattering particles in Brownian motion has been recently studied by Balog \textit{et al.}
\cite{balog06} for incident light in the coherent state, and the result following for this case from Eq.\ (\ref{db}) has been confirmed experimentally.

In conclusion, we developed a quantum description of dynamic multiple light scattering in fluctuating disordered media.
We calculated the fluctuation and the autocorrelation function of photon number operator in transmission through
a slab of mobile scattering particles. We applied our results to estimate the effect of
disorder on the information capacity of a quantum communication channel operating in a disordered environment and to
discuss the possibilities of performing diffusing-wave spectroscopy with squeezed light. Our results suggest that
using nonclassical light (single photons, light in number states, squeezed light) opens interesting perspectives
in optics of disordered media.

The author acknowledges fruitful discussions with R. Maynard and F. Scheffold. This work is supported by the French ANR and the French Ministry of Education and Research.



\begin{thebibliography}{99}
\bibitem{fnote}
The generalized continuum number state with an envelope $\xi(t)$ is constructed from the vacuum state $| 0 \rangle$ according to
$| n, \xi \rangle = (n!)^{-1/2} [{\hat A}^{\dagger}]^{n} | 0 \rangle$, where
${\hat A}^{\dagger} = \int \mathrm{d}t \xi(t) {\hat a}^{\dagger}(t)$ \cite{hussain92,blow90}.
\end{thebibliography}
\end{document}